\documentclass[aip,jcp,reprint]{revtex4-1}
\renewcommand{\vec}[1]{\mbox{\boldmath$#1$}}

\newcommand{\me}{\mathrm{e}}

\newcommand{\dif}{\mathrm{d}}

\usepackage{graphicx}

\usepackage{amsmath}

\usepackage{amsfonts}

\usepackage{dcolumn}

\usepackage{bm}

\begin{document}

\title{Direct Numerical Test of the Statistical Mechanical Theory of Hydrophobic Interactions}
\author{M. I. Chaudhari}\email{chaudhari.84@gmail.com}
\author{S. Holleran}\email{shollera@tulane.edu}
\author{H. S. Ashbaugh}\email{hanka@tulane.edu}
\author{L. R. Pratt}\email{lpratt@tulane.edu}
\affiliation{Department of Chemical and Biomolecular Engineering, Tulane
University, New Orleans, LA 70118}

             
\begin{abstract} 

This work tests the statistical mechanical theory of hydrophobic
interactions, isolates consequences of excluded volume interactions, and obtains $B_2$ for those purposes.
Cavity methods that are particularly appropriate for
study of hydrophobic interactions between atomic-size hard spheres in
liquid water are developed and applied to test aspects of the
Pratt-Chandler (PC) theory that have not been tested. Contact
hydrophobic interactions between Ar-size hard-spheres in water are
significantly more attractive than predicted by the PC theory. The
corresponding results for the osmotic second virial coefficient are
attractive ($B_2 <0$), and  more attractive with increasing temperature
($\Delta B_2/\Delta T < 0$) in the temperature range 300~K$\le T
\le$360~K. This information has not been available previously, but is
essential for development of the molecular-scale statistical
mechanical theory of hydrophobic interactions, particularly
for better definition of the role of attractive intermolecular interactions
associated with the solutes.

\end{abstract}

\maketitle

\section{Introduction}
Hydrophobic interactions are universally acknowledged as fundamental
contributions to the stability of folded or aggregated biomolecular
structures in water.  But hydrophobic interactions are also expected to
become more favorable with increasing temperature for physiological
temperatures.  On this basis, hydrophobic interactions are principally
entropic interactions.

Hydrophobic interactions can then be described as favorable for
aggregation \emph{and endothermic} at moderate temperatures. The osmotic
second virial coefficient 
\begin{multline} B_2 = \lim_{R \to
\infty}\left\{-2\pi \int_0^R \left\lbrack g_{\mathrm{AA}}\left(r\right)
-1 \right\rbrack r^2\dif r\right\} \\
\equiv \lim_{R \to
\infty}B_2\left\lbrack R\right\rbrack 
\label{eq:B2} 
\end{multline} 
is the solution thermodynamic metric for assessment of attractive and
repulsive character of hydrophobic interactions between an AA solute
pair.  Here $g_{\mathrm{AA}}\left(r\right)$  is the usual radial
distribution function of AA pairs at infinite dilution. Positive values
of $B_2$ raise the osmotic pressure, and indicate preponderance of
repulsive effects. Negative values of $B_2$ lower the osmotic pressure
and characterize interactions that are attractive on balance.

The entropic aspect of hydrophobic interactions requires statistical
thermodynamics for explanation.  Since the molecular theory of
hydrophobic interactions thus requires specified intermolecular
interactions and defensible statistical mechanics, the theory of
hydrophobic interactions has been intermittent and only partially
successful. The approximate Pratt-Chandler (PC) theory
\cite{PRATTLR:Thehe} was the first  prediction of molecular-scale
$g_{\mathrm{AA}}\left(r\right)$ for inert atom solutes in water and thus
the first molecular-scale prediction of $B_2$. Though the PC theory
predicted attractive and repulsive features in unprecedented detail, it
did not straightforwardly conform to the expectation that hydrophobic
interactions as expressed by $B_2$ should be \emph{attractive and
endothermic.} The PC theory was immediately controversial.\cite{Chan:79,Rossky:1980vh} Experiments for benzene, and
perfluorobenzene \cite{TUcker:1979wk,BernalP:VAPSOH} disagreed with the
PC theory for atomic-size hard-sphere  solutes. Explanations for the
discrepancy were suggested
\cite{PrattLR:HYDIAO,PrattLR:Hydsns,PrattLR:Effsaf} but the underlying
controversy has persisted.

One challenge for addressing this controversy is that the integrated
quantity $B_2$, and particularly the $R\to\infty$ limit, is a subtle
target for molecular simulation calculations. Another challenge is that
this problem requires analysis of temperature ($T$) dependences in a
limited $T$ range.

\begin{figure}[h] \includegraphics[width=3.0in]{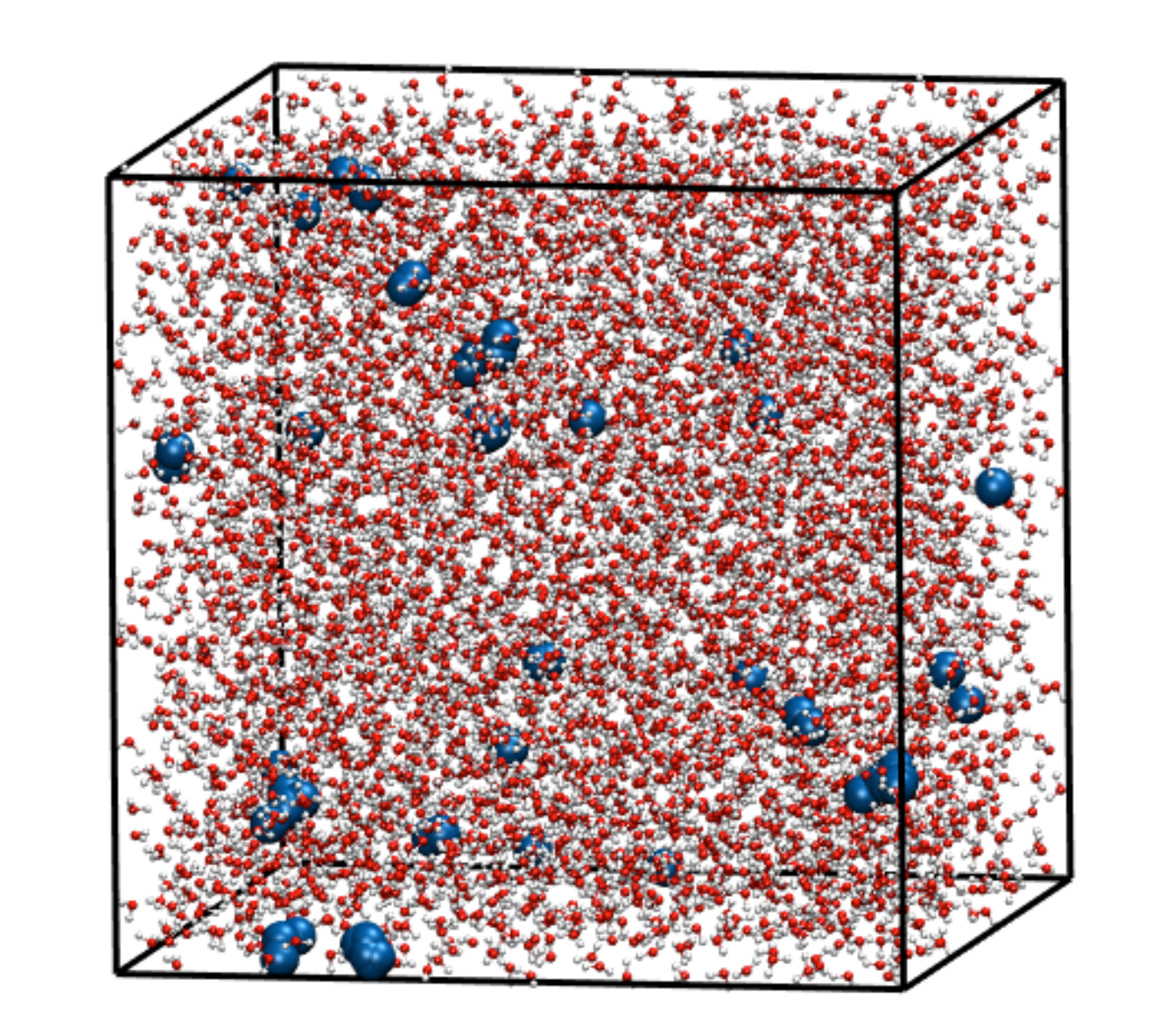} \caption{A
configuration of 5$\times 10^3$ water molecules (red and silver)
together with the inclusions (blue spheres) identified by $n_\mathrm{t}$
= 2$\times 10^5$ trial placements of a hard sphere with distance of
closest approach to an oxygen atom of 0.31~nm. This  size corresponds
approximately to an Ar solute for which the van der Waals length
parameter $\sigma_\mathrm{A}$ is about  0.34~nm,\cite{lowHill} thus
adopting 0.31~nm - 0.17~nm = 0.14~nm as a van der Waals contact radius
of the water oxygen atom.   Hard sphere solutes of this size are about
maximal for water-oxygen contact density.\cite{AshbaughHS:Colspt}
}\label{afoto}
\end{figure}

But further challenges remain, and a summary of the substantial efforts
to resolve this issue was given recently.\cite{Asthagiri:2008in}  The
broad conclusion is that simulation tests of the theory have differed
enough in details that a conclusive test of the PC theory for
hydrophobic interactions has been elusive. For example, $B_2$'s for the
specific cases of  Ar and CH$_4$ solutes have been estimated
\cite{WATANABEK:MOLSOT} to be remarkably small,  due evidently to
substantial cancellation between repulsive and attractive force effects.
In the same context,\cite{WATANABEK:MOLSOT} $B_2$ for Kr(aq) has been
estimated to be repulsive (positive).   Clearly, attractive and
repulsive interactions can play conflicting roles with the consequence
that merely realistic simulation of a case of interest, \emph{e.g.,}
CH$_4$, might not provide helpful physical conclusions. Indeed,
the case of atomic-size hard-sphere solutes has not been treated
specifically because hard-sphere models are inconvenient within common
molecular dynamics tools.

As another example, solute non-pairwise interactions associated with
polarizabilities were of simulation interest for a time.\cite{PrattLR:Molthe} Those interests have waned inconclusively, \cite{YoungWS:Arhe,PrattLR:Molthe} undoubtedly because of a lack of a molecular
theory that could sharpen the questions and make the answers more
permanent.  Such differences are indeed broadly expected to be details,
but in this context those details have been features of controversy.

In parallel with simulation efforts, the foundation of the molecular
theory of hydrophobic effects has undergone a surprizing renovation.\cite{PrattLR:Molthe,AshbaughHS:Colspt} The renovated theory directly
exploits molecular simulation data and substantially amends the PC
theory, but gives unexpected support for some of the conceptual
ingredients of that theory.\cite{PrattLR:Molthe} For  example, the
concern that the PC theory neglected molecular orientational structure
of liquid water is now generally recognized as not well founded. The
same can be said about the scaled-particle approaches applicable to the
present problem.\cite{AshbaughHS:Colspt,Stillinger:73}  The renovation
of the molecular theory is mostly due to aggressive extension of the
scaled-particle-theory concepts.   Indeed, the success of the
scaled-particle theory extensions
\cite{AshbaughHS:Colspt,Ashbaugh:2007bd} relieves the most serious
objections to the PC theory which can then be seen as merely an
approximate theory of liquid solutions.  Thus applied to molecular-scale
problems, the PC theory has about the same conceptual status as the
distinct minimal scaled-particle theories.  Of course, the PC theory
(and the information theory approach \cite{Hummer:PNAS:96} which is a
hybrid) addresses hydrophobic interactions which have not been addressed
by scaled-particle approaches.

Here we address the issue of $B_2$ from the renovated point-of-view. We
find that  $B_2$ for atomic-size hard spheres in water is attractive
($B_2 <0$), and more attractive with increasing temperature ($\Delta
B_2/\Delta T < 0$).

\section{Theory} We seek the \emph{cavity correlation function}
\begin{eqnarray} y_{\mathrm{AA}}(r) = \exp{\left\lbrack
u_{\mathrm{AA}}(r)/k_\mathrm{B}T\right\rbrack} g_{\mathrm{AA}}(r)
\end{eqnarray} for atomic-size hard spheres on the basis of the
potential distribution theorem (or \emph{test particle}) approach
\cite{BPP,arcc2012} \begin{eqnarray} y_{\mathrm{AA}}(1,2) =
\frac{\left\langle \me^{-\beta \Delta
U_\mathrm{AA}^{(2)}}\left.\right\vert 1,2\right\rangle_0} {\left\langle
\me^{-\beta \Delta U_\mathrm{A}^{(1)}}\right\rangle_0\left\langle
\me^{-\beta \Delta U_\mathrm{A}^{(1)}}\right\rangle_0} ~.
\label{eq:g-pdt} \end{eqnarray} This formula is cast for evaluation on
the basis of trial placements of hard-spheres at specific points, here
the two points $\left(1,2\right)$. Of course, we expect
$y_{\mathrm{AA}}(1,2) = y_{\mathrm{AA}}(r)$ to depend only on the
magnitude $r$ of the displacement between positions 1 and 2. The
notation $\left\langle \ldots \right\rangle_0$ indicates the average
over the configurations of water without the solutes present. $\Delta
U_\mathrm{A}^{(1)} = U(N+1) - U(N) - U(1)$ is the binding energy for
insertion of an A atom, and for the hard sphere case considered here is
either zero (no overlap with a water oxygen atom) or positive infinity.
Thus $\me^{-\beta \Delta U_\mathrm{A}^{(1)}}$ is an indicator function
for  permissibility of an insertion at a point considered. We will treat
the case that $\Delta U_\mathrm{AA}^{(2)}$ for two trial placements is
additive, $\Delta U_\mathrm{A}^{(1)} + \Delta U_\mathrm{A}^{(1)}$ for
the AA atoms considered.

We rearrange this formula to make the numerical estimation transparent,
and this rearrangement overlaps a recent discussion of the
Kirkwood-Salzburg theory.\cite{BPP} Note that the denominator factors of
Eq.~\eqref{eq:g-pdt}, being averages of indicator functions, are
probabilities.  Following the primitive understanding of conditional
probabilities $p(A \vert B) = p(A B)/p(B),$ we use one of those
denominator probabilities to introduce the expectation
\emph{conditional} on permissibility of that initial insertion. Taking
the position of that first insertion to be the origin
$\vec{0}$ we write 
\begin{eqnarray} y_{\mathrm{AA}}(r) =
\frac{\left\langle \me^{-\beta \Delta
U_\mathrm{A}^{(1)}}\left.\right\vert \vec{r}\right\rangle_0}
{\left\langle \me^{-\beta \Delta U_\mathrm{A}^{(1)}}\right\rangle_0} ~,
\label{eq-yconditional} \end{eqnarray} 
where $\vec{r}$ is the position of a trial placement relative to a
permissible insertion.  The average indicated in the numerator is
conditional on the permissible placement at $\vec{0}$, though we do not
set-up a further notation for that. Eq.~\eqref{eq-yconditional}
expresses the well-known zero-separation theorem
\cite{Hoover:1962jc,Meeron:1968jn,Devore:1979cm} \begin{eqnarray}
y_{\mathrm{AA}}(0) = \frac{1} {\left\langle \me^{-\beta \Delta
U_\mathrm{A}^{(1)}}\right\rangle_0} ~, \label{eq-zerosepartion}
\end{eqnarray} since the numerator is one (1) under the condition of a
permissible insertion at $\vec{0}$.

To estimate the ratio  (Eq.~\eqref{eq-yconditional}) we exploit
\emph{many} ($n_\mathrm{t}$) trial placements into the system volume
$V$, for each one of $n_\mathrm{c}$ configurations, $c$
(Fig.~\ref{afoto}). Those trial points will have the density
$n_\mathrm{t}/V$ and are statistically uniform.   Out of the
$n_\mathrm{t}$ trial points, a smaller number $n_\mathrm{s}(c)$ are
permissible placements, and we estimate the denominator of
Eq.~\eqref{eq-yconditional} with $\sum_c n_\mathrm{s}(c)/
\left(n_\mathrm{c}n_\mathrm{t}\right) =  \overline{n}_\mathrm{s}/
n_\mathrm{t}$.

\begin{figure}[h] \includegraphics[width=3.0in]{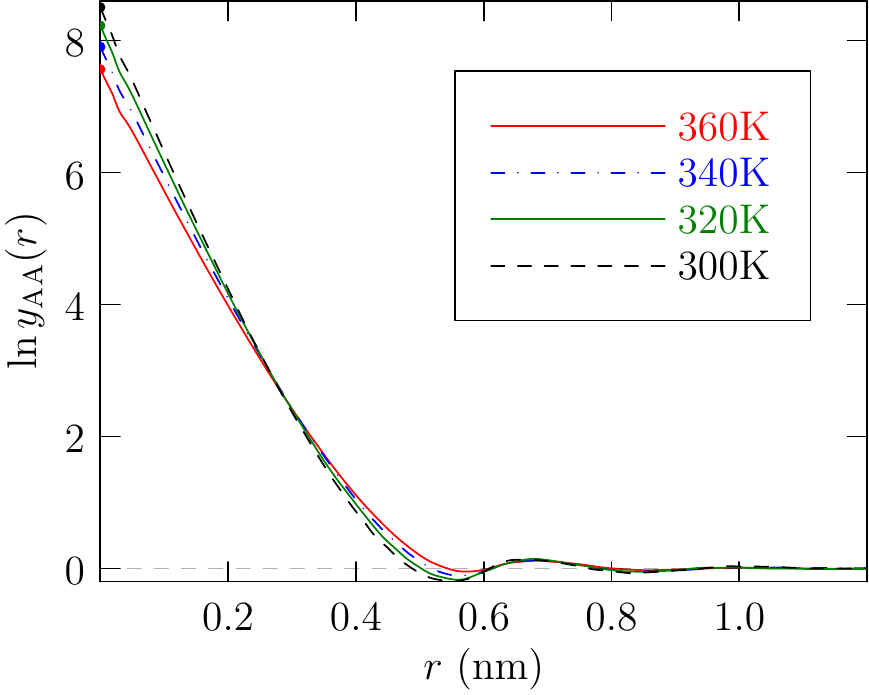}
\caption{Cavity correlation functions for hard-sphere solutes ion liquid
water at constant pressure $p$ = 1~atm, and four different temperatures.
The spheres have van der Waals radius of 0.17~nm and
distance-of-closest-approach to a water oxygen atom of 0.31~nm.  The
dots on the left vertical axis are values calculated independently on
the basis of the zero-separation theorem.}\label{fig:lnyAASmooth}
\end{figure}

We expect $\left(n_\mathrm{t}-1\right)\Delta V /V$ of those trial
placements to land in a volume $\Delta V$ which is a thin shell of
radius $r>0$ surrounding a permissible insertion. Let's denote by
$\Delta n_\mathrm{s}(r;c)$ the number of permissible placements obtained
in the shell for configuration $c$.   We estimate the numerator of
Eq.~\eqref{eq-yconditional} as \begin{eqnarray} \frac{\sum_c\Delta
n_\mathrm{s}(r;c)}{\left(n_\mathrm{t}-1\right)n_\mathrm{c}\Delta V /V} =
\frac{\Delta
\overline{n}_\mathrm{s}(r)}{\left(n_\mathrm{t}-1\right)\Delta V /V}~.
\end{eqnarray} Combining these results, we have the estimate
\begin{eqnarray} \left(\frac{ \overline{n}_\mathrm{s}}{V}
\right)y_{\mathrm{AA}}(r)\Delta V = \Delta  \overline{n}_\mathrm{s}(r)~,
\label{eq-yest} \end{eqnarray} when $n_\mathrm{t} \rightarrow \infty$.
This is the same formula as if the permissible insertions
were actual particles though that is not the case here.

The formula Eq.~\eqref{eq-yest} is operationally consistent with the
zero-separation theorem Eq.~\eqref{eq-zerosepartion} according to the
following argument. Consider a small volume element surrounding the
position $\vec{r} = \vec{0}$ that is known to be a permissible
placement.  We expect that all trial placements in this region should be
permissible so $\Delta \overline{n}_\mathrm{s}(\vec{0})\approx
n_\mathrm{t}\Delta V /V$ on the right-side of Eq.~\eqref{eq-yest}.
Therefore, $y_{\mathrm{AA}}(0)\approx
n_\mathrm{t}/\overline{n}_\mathrm{s}$ which is the operational content
of Eq.~\eqref{eq-zerosepartion}.

\section{Results and Discussion} Using standard methods (detailed
below), this approach was implemented for the case that the hard-sphere
distance-of-closest-approach to an oxygen atom was 0.31~nm,
corresponding approximately to the case of an Ar solute.   Larger
solutes would make the present calculations prohibitively difficult. The
results for  $\ln y_{\mathrm{AA}}(r)$ (Fig.~\ref{fig:lnyAASmooth})
operationally satisfy the zero-separation theorem, show strong
hydrophobic attraction as short-distances, and solvent-separated
hydrophobic attraction qualitatively consistent with the PC theory.

\begin{figure}[h] \includegraphics[width=3.0in]{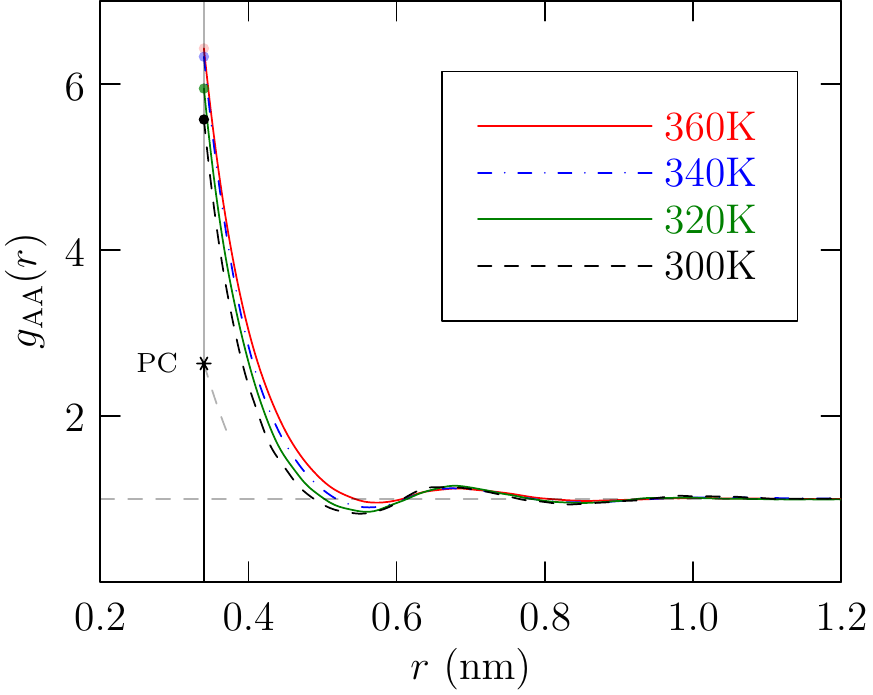}
\caption{Radial distribution functions for hard-sphere solutes in liquid
water at constant pressure $p$ = 1~atm, and four different temperatures.
The spheres have van der Waals radius of 0.17~nm and
distance-of-closest-approach to a water oxygen atom of 0.31~nm. The
prediction of the PC theory \cite{PRATTLR:Thehe} at $T=300$~K for the
contact value, $g_{\mathrm{AA}}(r$ = 0.34~nm), is shown by the star. The
contact values determined here are consistent with the information
theory model \cite{Hummer:PNAS:96} result published previously, and more
than twice  larger than the predictions of the PC theory. These contact
values are higher for higher temperatures, indicating stronger
hydrophobic contact attractions at higher temperatures.
}\label{fig:gAASmooth}
\end{figure}

The radial distribution function $g_{\mathrm{AA}}(r) =
y_{\mathrm{AA}}(r)$ for  $r\ge  2\times$ 0.17~nm = 0.34~nm (and zero
otherwise, Fig.~\ref{fig:gAASmooth}), scrutinizes these properties
more closely.  The contact values, $g_{\mathrm{AA}}(r$ = 0.34~nm),
determined here are consistent with the information theory model  result
published previously,\cite{Hummer:PNAS:96} and more than twice  larger
than the predictions of the PC theory.  The contact values are higher
for higher temperatures, indicating stronger hydrophobic contact
attractions at higher temperatures.  The contact values of the PC theory
also increase with increasing $T$  \cite{PRATTLR:Thehe} but those
increases are small, and the PC contact values are sufficiently
different from the numerical values found here that the small increases
are not interesting.

It is commonplace for simulation calculations to determine the
temperature dependences of the AA potentials of mean force
$w_{\mathrm{AA}}(r) \equiv -k_\mathrm{B}T \ln g_{\mathrm{AA}}(r)$. Some
examples are
\cite{Smith:1992p14802,SMITHDE:Freeea,SKIPPERNT:Comsms,Ludemann:1996tg,%
LudemannS:Thetha,PaschekD:Temdhh,PaschekD:Heacea,Sobolewski:2009kq}. We
note below that the recent work of that type supports our conclusions
here (Fig.~\ref{fig:gAASmooth}).   But that helpful recent work also
addresses our present problems only indirectly because it does not test
the statistical mechanical theory, nor isolate aspects of interactions
of different physical type, nor obtain $B_2$ for those purposes. Based
on the experience,  the measurable osmotic second virial coefficients
are more subtle, and despite interesting suggestions,\cite{chaudhari_communication:_2010} $w_{\mathrm{AA}}(r)$'s have not
been measured for molecular-scale hydrophobic solutes.

\begin{figure}[h] \includegraphics[width=3.0in]{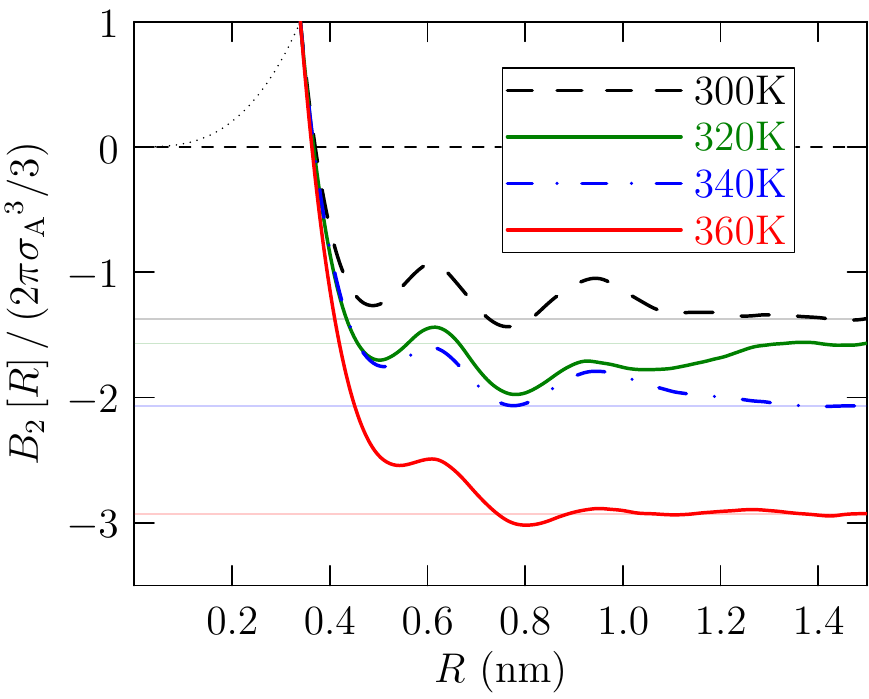}
\caption{Running integral $B_2 [R ]$ (Eq.~\eqref{eq:B2}) for assessment
of the $R \rightarrow \infty$ value.   The dotted curve for $R\le
0.34$~nm is the positive contribution from the hard-core of
$g_{\mathrm{AA}}\left(r\right)$ that is common to all here. The
suggested $R \rightarrow \infty$ values are decidedly negative
(attractive) and become more attractive at higher temperatures.  The
biggest negative contribution is associated with contact hydrophobic
attractions.  Solvent-separated hydrophobic attractions near $R\approx
0.7$~nm are distinct but smaller.  The predictions 
of the PC theory for $B_2$ in these circumstances are 
repulsive (positive).}\label{fig:b2r} \end{figure}

The $B_2$ integrals (Fig.~\ref{fig:b2r}) provide the solution
thermodynamic assessment of these distribution functions. The suggested
$B_2$ values are decidedly attractive (negative) for $T = 300$~K and
become more attractive at higher temperatures.  The biggest negative
contribution is associated with contact hydrophobic attractions.
Solvent-separated hydrophobic attractions near $R\approx 0.7$~nm are
distinct but smaller than contact hydrophobic interactions.

The significance of the solvent-separated hydrophobic interactions has
been much discussed following the ground-breaking  work of Pangali,
\emph{et al.} \cite{PANGALIC:AMCs,PANGALIC:Hydhap}. Those simulations
treated Lennard-Jones (LJ) model solutes somewhat similar to Kr or Xe
solutes. Simulation results (with LJ attractions) were compared with PC
theory (not treating LJ attractions though modified for continuous
repulsive interactions).   In view of the introductory discussion,
further analysis of the role of attractive interactions should be helpful.   But
additionally, since the simulation did not determine the additive
constant to $w_{\mathrm{AA}}(r)$, the comparison proceeded after
matching the two results at their minimum values, a convenient choice
that has been sometimes followed.\cite{YoungWS:Arhe,Chaudhari2014}  
This comparison can give the impression that the non-matched
solvent-separated hydrophobic interactions are unusually variable or
significant.   If, for the purposes of comparison, the
$w_{\mathrm{AA}}(r)$ were matched at solvent-separated radii, then the
Pangali, \emph{et al.} results show stronger contact hydrophobic
attractions than does the PC theory, qualitatively in agreement with the
present work. Recent molecular dynamics simulations for
Xe(aq) or  CH$_4$(aq) pairs
\cite{PaschekD:Temdhh,PaschekD:Heacea,Sobolewski:2009kq} agree qualitatively with the present results
and thus support this conclusion. Still the role of longer-ranged
attractive interactions remains to be studied.

Inclusion of longer-ranged attractive interactions, \emph{i.e.,} London
dispersion interactions, can change these $B_2$ values and trends
depending on the balance of solute-solute and solute-water interactions.\cite{PrattLR:Effsaf,Asthagiri:2008in}  Conclusive information for the
case that more general interactions are \emph{absent}, as with this
work, is crucial to justifying further analysis of more general
interactions. The LMF theory \cite{WeeksJD:Conlsi} is one promising
suggestion for how to proceed with inclusion of longer-ranged
interactions, has commonalities with earlier intuitive proposals,\cite{PrattLR:Effsaf,Asthagiri:2008in} and deserves further
development.

The temperature dependence observed here has generally been considered
counter-intuitive, and is  sometimes referred to as an \emph{inverse}
temperature dependence.  An explanation why this behavior might be
considered counter-intuitive is the following \cite{stillinger1980}:
hydrophobic association is typically rationalized as clumping of inert
solution inclusions due to specific structuring of their hydration
shells.  It might be guessed that the specific structuring should be
more significant at lower temperatures, so perhaps the hydrophobic
association should be stronger at lower temperatures, perhaps even more
important yet in super-cooled water.

Hydration-shell structuring surely is an important factor in hydrophobic
interactions.  What this argument does not address is the distinctive
equation of state of liquid water. Some well recognized peculiarities
occur at higher-than-physiological temperatures;  for example the
compressibility minimum occurs at 46~C, under these low pressure
conditions.  The eventual statistical mechanical explanation
\cite{Garde:1996to} of the similarly counter-intuitive \emph{entropy
convergence}  hydrophobic phenomenon (at $T \approx$ 130~C) depended
firstly on proper involvement of the actual equation of state of liquid
water.\cite{AshbaughHS:Colspt}  Indeed the calculations here model that
specific equation of state also.

\section{Conclusions}
This work tests the statistical mechanical theory of hydrophobic
interactions, isolates consequences of excluded volume interactions, and obtains $B_2$ for those purposes. Contact hydrophobic
interactions between Ar-size hard-spheres in water are significantly
more attractive than predicted by the Pratt-Chandler theory. The
corresponding $B_2$ results for atomic-size hard spheres in water are
attractive ($B_2 <0$), and  more attractive with increasing temperature
($\Delta B_2/\Delta T < 0$) in the temperature range 300~K$\le T 
\le$360~K. This information is essential for further development of the
molecular-scale statistical mechanical theory of hydrophobic
interactions.

\section{Methods} 
The GROMACS \cite{Hess:2008db} package and the SPC/E \cite{BEREND87A}
model was used to simulate liquid water. This simulation adopted the
isothermal-isobaric ensemble at four different temperatures (\textit{T}=
300 K, 320 K, 340 K and 360 K). A Nose-Hoover thermostat maintained the
temperature and a Parinello-Rahman barostat was used to establish the
pressure at 1 atm. Bonds involving hydrogen atoms were constrained by
the LINCS algorithm. Conventional periodic boundary conditions and
particle mesh Ewald, with a real-space cut-off at 1 nm, was used to
treat long-range interactions. Simulation cells containing 5$\times
10^3$ randomly placed water molecules were created utilizing PACKMOL
\cite{Martinez:2009di} to  match the experimental density approximately.
 After 1$\times 10^4$ steps of energy minimization and 2 ns of density
equilibration, trajectories of 20 ns (sampled 1/ps) were obtained at
each temperature. Each simulation frame was analyzed for cavities based
on $n_\mathrm{t}$ = 2$\times 10^5$ trial placements with a distance of
closest approach to an oxygen atom of 0.31~nm. Successful placements can
be considered as hard spheres of radius 0.17~nm corresponding approximately
an Ar atom.   The cavity analysis is about an order-of-magnitude more
computational effort than the generation of the molecular dynamics
trajectories.

Considering the integrand of Eq.~\eqref{eq:B2},
$g_{\mathrm{AA}}\left(r\right) \sim 1$ in the thermodynamic limit. For a
fixed particle numbers in a finite system, that subtracted uncorrelated
feature is less than 1 and the correction is $O(V^{-1})$
\cite{Lebowitz:1963tn}, vanishing in the thermodynamic limit. In the
present work, we do not have fixed numbers of A particles.



%

\end{document}